\documentstyle[epsf,bibnorm]{lamuphys}

\newcommand{\lesssim}{{\raisebox{0.25ex}{$<$}\hspace*{-0.78em}\raisebox{-0.95ex}{$\sim$}\ }}
\newcommand{\gtrsim}{{\raisebox{0.25ex}{$>$}\hspace*{-0.78em}\raisebox{-0.95ex}{$\sim$}\ }}

\begin{document}
\title{Non-linear Transport in Quantum-Hall Smectics}

\author{A.H. MacDonald and Matthew P.A. Fisher}
\institute{Department of Physics, Indiana University, Bloomington, IN
47405-4202\\
Institute for Theoretical Physics, University of California at
Santa Barbara, Santa Barbara CA 93106-4030}

\maketitle

\begin{abstract}

Recent transport experiments have established that
two-dimensional electron systems with high-index partial Landau level
filling, $\nu^{*} =\nu - \lbrack \nu \rbrack$, have ground states
with broken orientational symmetry.  
In a mean-field theory, the broken symmetry state consists of 
electron stripes with local filling factor $\lbrack \nu \rbrack + 1 $, separated
by hole stripes with filling factor $\lbrack \nu \rbrack$.  
We have recently developed a theory of these states in which the 
electron stripes are treated as one-dimensional electron systems
coupled by interactions and described by using a Luttinger liquid model.
Among other things, this theory predicts non-linearities of opposite sign in easy
and hard direction resistivities.  In this article we briefly review our
theory, focusing on its predictions for the dependence of non-linear transport
exponents on the separation $d$ between the two-dimensional electron system
and a co-planar screening layer.

\end{abstract}

\section{Introduction}

Recent transport experiments \cite{lilly,du,shayegan} have established
that the resistivity of a two-dimensional electron system with weak disorder 
and valence orbital Landau level filling $\nu^*$ close to $1/2$ 
is anisotropic when the valence orbital Landau level index $N \ge 2$.
Apparently the ground state spontaneously breaks orientational
symmetry, a property believed to be associated with the uni-directional
charge-density-wave ground states predicted \cite{fogler,moessner}
in this regime by Hartree-Fock mean-field theory \cite{mftapology}.
The charge-density-wave (CDW) state consists of electron stripes
of width $a \nu^*$ with local Landau level filling factor
$\nu=\lbrack \nu \rbrack +1$, separated by hole stripes of width $a (1 - \nu^*)$ with local
filling factor $\nu = \lbrack \nu \rbrack$. Here $a$, the CDW period, is comparable
to the Landau level's cyclotron orbit diameter.  Because of
symmetry properties shared with the smectic state of 
classical liquid crystals, emphasized by Fradkin and Kivelson \cite{fradkin},
these states have been referred to as quantum Hall smectics, 
a practice we follow here.  The most important transport property of
quantum Hall smectics is that dissipation occurs over a
wide range of filling factors surrounding $\nu^* =1/2$ and
is not activated at low temperatures.
This behavior is {\em not} consistent with the properties of 
a CDW state, which would be pinned by the random disorder potential
and have a large gap for mobile quasiparticle excitations,
and suggests that, although Hartree-Fock theory hints at the energetic motivation for a
ground state with broken orientational symmetry, the description
which it provides of the ground state is flawed.

Several recent theoretical papers \cite{rezayi,fertig,phillips,jungwirth,wang,stern}
have addressed the properties of quantum Hall smectics and
the energetic competition between CDW states, compressible
composite-fermion states, and paired incompressible quantum Hall states.
In one recent paper \cite{usprb} we have described a theory of
quantum Hall smectics which starts from the Hartree-Fock theory 
ground state, recognizes the electron stripes as coupled
one-dimensional electron systems, and treats residual interaction
and disorder terms neglected by Hartree-Fock theory using
the convenient bosonization techniques of one-dimensional electron
physics.  The most important conclusions of this work are the
following: i) the quantum Hall smectic state is {\em never} the ground state
but instead is always unstable, for $\nu^*$ close to $1/2$ likely to an anisotropic electron
Wigner crystal state; ii) for $0.4 \lesssim \nu^* \lesssim 0.6$ the
interaction terms responsible for the Wigner crystal can be
neglected at temperatures available in a dilution fridge; and
iii) weak disorder which scatters electrons from stripe to stripe,
enabling hard-direction transport, leads to non-linear
transport. In this article we emphasize and expand on an experimentally
important prediction of this work, namely that the strength of
the transport non-linearity is sensitive to the nature of the
electron-electron interaction.  In particular we predict that the
transport non-linearity can be enhanced by placing a screening
plane close to the two-dimensional electron system.

In Section II we explain our theory of
transport in quantum Hall smectics and discuss how the
coefficients which govern the power-law behavior of the
differential resistivity for weak disorder are related to
correlations in the coupled one-dimensional electron stripes.
In Section III we briefly review our theory of these correlations
and explain why long-range electron-electron interactions
weaken transport non-linearities.  In Section IV we
present numerical results for the non-linear transport coefficients 
for a model in which interactions in
the two-dimensional electron layer are screened by a metallic
layer co-planar with the electron system but separated from it 
by a distance $d$.  We conclude in Section V with a brief summary.

\section{Anisotropic Transport Properties}

Our transport theory \cite{usprb} is built on a semiclassical Boltzmann-like
approach; microscopic physics enters only in the calculation
of scattering rates.  We choose a coordinate system where 
the $\hat x$ (horizontal) direction runs along the stripes which 
are separated in the $\hat y$ (vertical) direction.
We assume that the charge
density wave itself is pinned and immobilized by both the edges of the sample
and weak impurities which couple to the electrons within the stripes. In this
case, collective sliding motion of the charge-density will be absent, and 
electrical transport will be dominated by single-electron
scattering across and between electron stripes.
An important property of electronic states in the quantum Hall
regime, is the spatial separation of states which carry current in
opposite directions.  In the case of the electron stripes in
quantum Hall smectics, the Fermi edge states which carry
oppositely directed currents along the stripes (left-going and right-going) 
are located on opposite sides of the stripe.  
Translational invariance along the $\hat x$ direction, allows us
to use a Landau gauge where this component of wavevector $\hbar k_{x}$ is a
good quantum number in the absence of interactions and disorder.
The single-particle states at the stripe edges have velocity
magnitude $v_{F}$, the Fermi velocity.  In the Landau gauge, $\hat x$ direction momenta are
related to $\hat y$ direction positions by $k_{x} = y/\ell^2$ where
$\ell \equiv (\hbar c/eB)^{1/2}$ is the magnetic length, so that states 
on opposite sides of the same stripe differ in momenta by a $\nu^* a /\ell^2$
and the adjacent sides of neighboring stripes differ in momenta by $(1-\nu^*) a/\ell^2$.

We now summarize the basic assumptions on which our semiclassical
transport theory is 
based and quote the expressions implied by these assumptions.
For further details see Ref.~\cite{usprb}.
We assume that in the steady state, each edge of each stripe is
characterized by a local chemical potential.  Translational invariance 
in the $\hat y$ direction implies
that the chemical potential drops across each stripe and between
any two adjacent stripes are the same,
mf $\mu$ is the chemical potential drop across an electron stripe,
it follows that the potential drop between stripes is $ e E_{y} a -
\mu$, where $E_{y}$ is the hard-direction electric field, the field
which supports a steady state transport across the stripes.  The electric field
in the $\hat x$ direction produces a semiclassical drift in 
momentum space which drives the system from equilibrium.  We assume 
that disorder scattering across and between stripes, then attempts 
to reestablish equilibrium and that the drift and scattering 
processes are in balance in the steady state.  The scattering currents in the hard-direction
are characterized by relaxation times $\tau_{e}$ and $\tau_{h}$
respectively.  The current along a stripe is analogous to the quantized current
in a long narrow Hall bar and is proportional to the 
chemical potential difference across that stripe. 
Combining these ingredients leads \cite{usprb} to the following expressions for the
resistivities:
\begin{eqnarray}
\rho_{\rm easy} &=& \frac{h}{e^{2}} \, \frac{1}{\tau_{e} ( \lbrack \nu \rbrack
+1)^{2} + \tau_{h} \lbrack \nu \rbrack^2 } \, \frac{a}{v_{F}} \nonumber\\
\rho_{\rm hard} &=& \frac{h}{e^{2}} \, \frac{1}{\tau_{e} ( \lbrack \nu \rbrack
+1)^{2} + \tau_{h} \lbrack \nu \rbrack^2} \, \frac{v_{F} \tau_{e} \tau_{h}}{a}
\nonumber\\
\rho_{\rm hall} &=& \frac{h}{e^{2}} \frac{1}{\tau_{e} ( \lbrack \nu \rbrack +1)^{2}
+ \tau_{h} \lbrack \nu \rbrack^2} \, (\lbrack \nu \rbrack +1) \tau_{e} +
\lbrack \nu \rbrack \tau_{h},
\label{rho}
\end{eqnarray}
where $\rho_{\rm easy}=\rho_{xx}$, $\rho_{\rm hard}=\rho_{yy}$, and
$\rho_{\rm hall}=\rho_{xy}$.

For $\nu^* =1/2$, this theory makes a parameter free prediction
for the product $\rho_{easy} \rho_{hard}$ which has been
confirmed experimentally \cite{lilyprivate}.  In fact, as emphasized \cite{stern}
by van Oppen {\it et al.}, this feature of our results has a greater validity than
would be suggested by our assumption of largely intact electron stripes.
Our main interest here however, is in expanding on our predictions \cite{usprb} for
non-linearities in the easy and hard direction differential
resistivities.  These predictions were made on the basis of a
simple lowest order renormalization group scheme
for handling the infrared divergence which appear when
disorder terms which scatter electrons either across or between
stripes are treated perturbatively.  This analysis leads to
\begin{eqnarray}
\frac{1}{\tau_{e}} \equiv \Gamma_{e} &\sim& \Gamma_{e}^{(0)} (V_{y}/ E_{c})^{2 \Delta_{e}-2} \nonumber\\
\frac{1}{\tau_{h}} \equiv \Gamma_{h} &\sim& \Gamma_{h}^{(0)} (V_{y} /E_{c})^{2 \Delta_{h}-2}
\label{nonlintrans}
\end{eqnarray}
where $\Gamma_{e}^{(0)}$ and $\Gamma_{h}^{(0)}$ are 
Golden-rule scattering rates at the characteristic microscopic energy scale $E_{c}$,
Here $\Delta_{e}$ is \cite{usprb} the scaling dimension of the operator which scatters
an electron across a stripe, which we discuss in the next section, and 
$\Delta_{h}$ is the scaling dimension of the operator which scatters 
an electron between neighboring stripes.
The values of $\Delta_{e}$ and $\Delta_{h}$ depend on correlations induced by 
electron-electron interaction between stripes, and are sensitive in particular to
the range of the microscopic electron-electron interaction.
At $\nu^* = 1/2$, the case on which we will concentrate, $\Delta_{e}= \Delta_{h}$. 

Given these expressions, it follows \cite{usprb} that  
the non-linear differential resistivity in the hard direction 
\begin{equation}
\frac{\partial V_{y}}{\partial I_{y}} \sim I_{y}^\alpha ,
\label{diffresisty}
\end{equation}
with an exponent $\alpha = 2(1-\Delta_{e})/(2\Delta_{e} -1)$.
Similar considerations apply for the easy direction 
current: 
\begin{equation}
\frac{\partial V_{x}}{\partial I_{x}} \sim I_{x}^\beta
\label{diffresistx}
\end{equation}
with an exponent $\beta = 2(\Delta_{e} -1)$.  In the next section we show that
both $\alpha$ and $\beta$ increase when the distance to the  
screening plane is comparable to or smaller than the CDW period.

\section{Quantum Smectic Model}

The CDW state of Hartree-Fock theory \cite{fogler,moessner} is a
single-Slater-determinant.  In the valence Landau level, groups of Landau-gauge 
single-particle states with adjacent $k_{x}$ (adjacent $\hat y$)
are occupied to form stripes and separated by groups which are unoccupied.
Small fluctuations in the positions and shapes of the stripes can be described
in terms of particle-hole excitations near the stripe edges. The residual
electron-electron interaction terms which scatter into these low
energy states are ignored in Hartree-Fock theory and 
fall into two classes: ``forward" scattering interactions which
conserve the number of electrons on each edge of every stripe, and ``backward"
scattering processes which do not. The latter processes involve large momentum
transfer and are unimportant \cite{usprb} at accessible temperatures for $\nu^*$
near $1/2$. The quantum smectic model \cite{usprb}, briefly
described in this section includes forward scattering only.  The
interactions are bilinear in the 1D electron density contributions from 
a particular edge of a particular stripe: $\rho_{n\alpha}(x)$, with $\alpha=\pm$.
Since the density of a single filled Landau level is 
$(2 \pi \ell^2)^{-1}$, the displacement of an edge
is related to its associated charge density contribution by
$u_{n\alpha}(x) = \alpha 2 \pi \ell^{2} \rho_{n\alpha}(x)$.
The quadratic Hamiltonian
which describes the {\it classical} energetics for small fluctuations
has the following general form:
\begin{eqnarray}
H_{0} &=& \frac{1}{2\ell^{2}} \int_{x,x'} \sum_{n,n'} u_{n\alpha}(x) D_{\alpha
\beta}(x-x';n-n') u_{n'\beta}(x') \nonumber \\
&=& \frac{1}{2\ell^{2}} \int_{{\bf q}} u_{\alpha}(- {\bf q}) D_{\alpha
\beta}({\bf q}) u_{\beta}({\bf q}) ,
\label{hamiltonian}
\end{eqnarray}
where $\int_{{\bf q}} \equiv \int d^{2}{{\bf q}}/(2 \pi)^{2}$. Here the
$q_{y}$ integral is over the interval $(-\pi/a,\pi/a)$ and a high momentum
cutoff $\Lambda \sim 1/\ell$ is implicit for $q_{x}$.

Symmetry considerations constrain the form of the elastic kernel. In position
space, the kernel must be real and symmetric so that $D_{\alpha \beta}({\bf q})
= D^{*}_{\alpha \beta} (-{\bf q}) = D^{*}_{\beta \alpha}({\bf q})$. This
implies $D_{-+}({\bf q})=D^{*}_{+-}({\bf q})$ and ${\rm Im} D_{\alpha
\alpha}({\bf q}) = 0$. Parity invariance (under $x,n,+ \leftrightarrow
-x,-n,-$), implies moreover $D_{++}({\bf q})=D_{--}({\bf q})$. Thus, the
elastic kernel is fully specified by one real function, $D_{++}({\bf q})$, and
one complex function, $D_{+-}({\bf q})$.  
It will be important for our present interest that the 
Hamiltonian must be invariant under: $u_{n\alpha}(x) \rightarrow u_{n\alpha}(x)+
const$.  For short-range interactions this implies that at long wavelengths
\begin{equation}
D(q_{x}=0 ,q_{y}) =  K_{y} q_{y}^{2} + ....  ,
\end{equation}
characteristic of classical smectic elasticity.  As we will discuss, this 
conclusion must be modified in the case of long-range interactions.

A {\it quantum} theory of the Quantum-Hall smectic \cite{usprb} is obtained
by imposing Kac-Moody commutation relations on the chiral densities:
\begin{equation}
[\rho_{n\alpha}(x),\rho_{n'\beta}(x')] = \frac{i}{2\pi} \alpha
\delta_{\alpha,\beta} \delta_{n,n'} \partial_{x} \delta(x-x').
\end{equation}
Together with Eq.(~\ref{hamiltonian}), this relationship 
fully specifies the quantum dynamics.  Electron
operators in the chiral edge modes are related to the 1D densities via the usual
bosonic phase fields: $\psi_{n\alpha} \sim e^{i\phi_{n\alpha}}$ with
$\rho_{n\alpha} = \alpha \partial_{x} \phi_{n\alpha} /2\pi$.

Quantum properties of the smectic can be computed from the imaginary-time
action,
\begin{eqnarray}
S_{0} &=& \int_{x,\tau} \frac{1}{4\pi} \sum_{n,\alpha} i \alpha \partial_{\tau}
\phi_{n,\alpha} \, \partial_{x} \phi_{n,\alpha} + \int_{\tau} H_{0} \nonumber \\
&=& \frac{1}{2} \int_{{\bf q},\omega} \phi_{\alpha}(-{\bf q},-\omega)
M_{\alpha,\beta} ({\bf q},\omega) \phi_{\beta}({\bf q},\omega) ,
\label{action}
\end{eqnarray}
where in an obvious matrix notation,
\begin{equation}
{\bf M}({\bf q},\omega) = (i \omega q_{x}/2\pi) {\bf\sigma}^{z} + (q_{x}
\ell)^{2} {\bf D}({\bf q}).
\label{actionkernel}
\end{equation}
Correlation functions follow from Wick's theorem and the momentum space
correlator $\langle \phi_{\alpha} \phi_{\beta} \rangle = {\bf M}^{-1}$ with
\begin{equation}
{\bf M}^{-1}({\bf q},\omega) = {\bf\sigma}_{z} {\bf M}({\bf q},-\omega)
{\bf\sigma}_{z} / {\rm det} {\bf M}({\bf q},\omega) .
\end{equation}

The effect of weak disorder on transport in quantum Hall smectics
depends sensitively on the elastic constants {\it at} $q_{x}=0$.
In this limit the relevant excited states are simply Slater determinants with straight
stripe edges displaced from those of the Hartree-Fock theory ground state. By
evaluating the expectation value of the microscopic Hamiltonian in a state with
arbitrary stripe edge locations we find that
\begin{equation}
D_{\alpha \beta}(q_{x}=0,q_{y}) = \delta_{\alpha \beta} D_{0} +
\alpha \beta \frac{a}{4 \pi^2 \ell^2}
\sum_{n} e^{iq_{y}an} \Gamma(y^{0}_{n\alpha} - y^{0}_{0\beta}) ,
\label{elastic}
\end{equation}
where the constant $D_{0}$ is such that the condition $\sum_{\alpha \beta}
D_{\alpha \beta}({\bf q}=0) =0$, and the positions $y^{0}_{n\pm} = a (n \pm \nu^{*}/2)$
are the ground state stripe edge locations.  Here, $\Gamma(y)$ is the
interaction potential between two electrons located in guiding center states a
distance $y$ apart:
\begin{equation}
\Gamma(y) = U(0,y/\ell^{2}) - U(y/\ell^{2},0),
\label{gamma}
\end{equation}
\begin{equation}
U(q,k) = \int \frac{d p}{2 \pi} e^{- (q^{2}+p^{2})\ell^{2}/2} \, V_{\rm
eff}^{N}(q,p) e^{- i p k \ell^{2}}.
\label{matrixelement}
\end{equation}
The two terms in Eq.~(\ref{gamma}) are direct and exchange contributions. In
Eq.~(\ref{matrixelement}), $V_{\rm eff}^{N}(q,p)$ is the Fourier transform of
the effective 2D electron interaction which incorporates form-factors
\cite{jungwirth} dependent on the Landau level index. 
$N$ and the ground subband wavefunction
of the host semiconductor heterojunction or quantum well.
The smectic states have relatively long periods proportional to the
cyclotron orbit radii.
Explicit calculations \cite{fogler,phillips,jungwirth} show
that $a \gtrsim 6 \ell$ for $N =2$.  It follows that the
exchange contribution to $\Gamma(y)$ is small and that
$\Gamma(y)$ decreases with stripe separation in the relevant range.
In this paper we address the influence of a metallic screening 
plane which cuts off this interaction at large distances.  
For $y \lesssim d$, $\Gamma(y) \sim 2 e^2 \ln (d/y)$, decreasing 
extremely slowly with $y$.
For separation $y$ larger than the distance $d$ to the screening plane,
$\Gamma(y) \sim y^{-2}$, making the sum over $n$ in Eq.~\ref{elastic} convergent.

\section{Screening Dependence of Scaling Dimensions}

The scaling dimension, $\Delta_{e}$, of the operator $e^{i(\phi_{n,+} - \phi_{n,-})}$ which 
scatters an electron across the $n$-th stripe is readily evaluated from
Eq.(~\ref{actionkernel}).  We find that 
\begin{equation}
\Delta_{e} = \int_{-\pi}^{\pi} { {d(qa) } \over {2\pi} } W(q_{x}=0,q)   .
\label{integral}
\end{equation}
Here, $W$ is the weight function,
\begin{equation}
W({\bf q}) = \frac{[D_{++}({\bf q}) +
{\rm Re}D_{+-}({\bf q})]}{[D_{++}^{2}({\bf q}) - |D_{+-}({\bf q})|^{2} ]^{1/2}} .
\label{wght}
\end{equation}

For 1D non-interacting electrons $\Delta_{e} =1$, so that disorder is relevant and
eventually leads to localization.  As discussed below, $\Delta_{e} < 1$ for 
quantum Hall smectics.  Disorder is even more relevant than in the non-interacting
electron case.  Nevertheless, since the samples in which the quantum Hall
smectic is observed are of extremely high quality, there should be a wide 
range of temperature over which its effects can be treated perturbatively.  
If $\Delta_{e}=1$ both hard direction ($\alpha$) and easy direction ($\beta$) non-linear 
transport exponents vanish.
We see from Eq.(~\ref{wght}) that $\Delta_{e} = 1$ 
if the average value of $W({\bf q_{y}})$ is one.  To understand the 
dependence of $\Delta_{e}$ on screening, we have to understand the 
dependence of $W(q_{y})$ on both wavevector and $d$.  

Note that $W$ is smaller than one, increasing the relevance of disorder, when
$D_{++}$ and ${\rm Re}D_{+-}$ are opposite in sign and similar in magnitude.
For each $q_{y}$ in Eq.~\ref{integral} 
the weighting factors are like those which enter in the calculation
of the scaling dimension of the operator which describes 
backscattering from disorder in an isolated one-dimensional electron system.
In continuum 1D models, $D_{++}$ has a contribution, proportional to the Fermi 
velocity, from the band energy and a contribution proportional to the interaction
between electrons traveling in the same direction, while $ - D_{+-}$ has only
an interaction contribution.  For a continuum model, the effective interactions 
between electrons traveling in different directions is the same as that between
electrons traveling in the same directions.  When the interaction term is 
much larger than the band term $D_{++}$ and $D_{+-}$ are opposite in sign and 
nearly equal in magnitude and $W$ is very small.  This is what happens, for 
example, for a 1D electron system in which long-range makes the Coulomb interaction very
strong at long wavelengths.  When $W$ is small, the 1D electron system is 
very close \cite{schulz} to an electron Wigner crystal, and disorder is very strongly relevant.
On the other hand when $D_{+-}$ is much smaller than $D_{++}$ we have a situation
analogous to that in a very weakly interacting Fermion system, in which disorder 
is relevant but the resistivity is linear when disorder can be treated perturbatively.

With this in mind we turn to a discussion of the quantum smectic, limiting our 
attention to the case $\nu^* =1/2$. 
Useful insight comes from examining the value of $W$ at the 
end points of the integration interval,
$q_{y} a= 0$ and $q_{y} a = \pi$ where both $D_{++}$ and $D_{+-}$ are real.
For $q_{y}=0$, invariance under a uniform
translation of the smectic implies that $D_{++} + D_{+-} =0$, 
so that $W(q_{y}=0) = 0$.  When all the electron stripes move together, 
the energetics is precisely like that of a single 1D system.  In the 
quantum Hall regime, there is no band energy, only interaction contributions 
from electrons traveling in the same direction, which appear in 
$D_{++}$, and interaction contributions from electrons traveling in 
opposite directions, which appear in $D_{+-}$.  The absence of a band 
contribution means that, for $q_{y} = 0$, $W$ vanishes independent of
the interaction's strength or range.
When $q_{y}a=\pi$, on the other hand, one has
\begin{equation}
D_{+-}(q_{y}a =\pi)  = \sum_{n} \frac{(-1)^{n} a}{ 4 \pi^2 \ell^2}
[\Gamma(an + a(1-\nu^{*})) - \Gamma(an+a\nu^{*})] .
\end{equation}
We see that $D_{+-}(q_{y}a =\pi)$ vanishes because the interaction between
the top of one stripe and the bottom of the same stripe is identical
to its interaction with the bottom of the next stripe up.  
The interactions between oppositely directed electrons effectively cancel
out, and we obtain $W=1$, just as we would for a non-interacting 1D system.
Thus $\Delta_{e}$ is determined by the average over $q_{y}$ of a 
weighting factor which interpolates between that of a 1D electron 
Wigner crystal at $q_{y}a=0$ and that of a non-interacting 1D electron system
at $q_{y}a=\pi$.

The average value of $W$ is determined
by the rate at which $W$ goes from its $q_{y}=0$ limit to its 
$q_{y}=\pi$ limit.  To obtain insight into what controls this, we 
consider first the limit of short-range interactions.  Since 
$\Gamma(y=0)$ vanishes due to the cancellations of its direct and 
exchange contributions, the short-range limit is obtained by 
taking only $\Gamma(a/2) \equiv \Gamma^* \ne 0$.  In this case,
it follows from Eq.(~\ref{elastic}) that 
$ (4 \pi^2 \ell^2) D_{+-}(q)/a =  - \Gamma^* (1 +\exp(-iqa))$ and 
that $(4 \pi^2 \ell^2 D_{++}(q)/a = 2 \Gamma^*$, and therefore that 
$W(q) = |sin(qa/2)|$.  The numerator of Eq.(~\ref{wght}) vanishes
like $q^2$ for $q \to 0$, and the denominator, which is proportional
to the collective mode velocity, vanishes like $|q|$.  Note that 
the expression for $W(q)$ is independent of $\Gamma^*$.  The average 
of $W(q)$ may be evaluated analytically in this case, and we obtain
for short-range interactions $\Delta_{e}=2/\pi \approx 0.6366$.

For the realistic case, analytic calculations are no longer possible,
but the behavior of $W$ can be simplified, at least at small $q$
if the exchange contribution to $\Gamma (y)$ is neglected in constructing
$D_{\alpha,\beta}(q)$.  In this case $\Gamma (y)$ is the simply the 
1D transform to coordinate space of the reciprocal space interaction
$U(p) \equiv \exp(-p^2 \ell^2/2) V_{eff}^{N}(0,p)$, i.e. it is the Coulomb
interaction between lines of charged smeared by $N$-dependent
cyclotron orbit form factors.  The components of $D_{\alpha,\beta}(q)$,
are then Fourier transforms back to reciprocal space, but with additional
`umklapp' terms because this transform is discrete.  We find that 
\begin{eqnarray}
D_{+-}(q) &=& \frac{-1}{4 \pi^2 \ell^2} \exp(-i qa/2) \sum_{j=-\infty}^{\infty} (-)^j
U(q+2 \pi j/a)  \nonumber \\
D_{++}(q) &=& D_0  + 
 \frac{1}{4 \pi^2 \ell^2}  \sum_{j=-\infty}^{\infty} 
U(q+2 \pi j/a)  
\label{noexchange}
\end{eqnarray}
and 
\begin{equation}
D_0 = \frac{1}{4 \pi^2 \ell^2}  \sum_{j=-\infty}^{\infty} 
U(2 \pi j/a)  [ (-)^{j} -1].
\label{dnotnoexchange}
\end{equation}
For the case of a Coulomb interaction,
$U(q) = [2 \pi e^2 (1 - \exp (-2 q d))] /q $ where $d$ is the distance to
a screening plane described by an image charge model.
For large or infinite $d$, the $j=0$ terms dominate the 
sums in Eqs.~(\ref{noexchange}). The above expression for $U(q)$ 
applies when $q \ell  \lesssim 1$,
and therefore is always valid for the $j=0$ terms in Eqs.(~\ref{noexchange}).
Note that both $D_{++}(q)$ and $D_{\pm}(q)$ are proportional to $d$ for
large $d$ and diverge for $d \to \infty$.  On the other hand 
$D_0$ remains finite for $d \to \infty$  because the $j=0$ term is excluded from this sum.  
We emphasize that the numerical calculations whose results are shown below include
the exchange contributions neglected in deriving Eqs.(~\ref{noexchange}), and 
important at any value of $a$ when $d$ is not large.

\begin{figure}
\vspace*{3.5cm}
\centerline{\epsfxsize=2in
     \epsffile{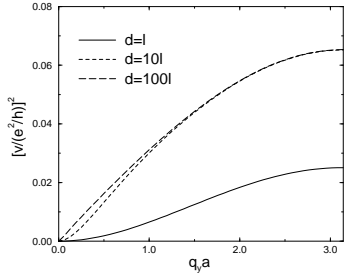}}
%\special{psfile=v2N2.eps}
\caption[]{\label{v2}
Square of the collective excitation velocity (in units of 
$e^2/(2 \pi \epsilon \hbar)$) in the $\hat x$ direction as a 
function of $q_{y}$ for $d=\ell$, $d = 10 \ell$ and $d = 100 \ell$.  These 
calculations are for $N=2$ and CDW period $a = 5.8 \ell$.  For the 
dielectric function of GaAs, this velocity unit has the value 
$ 4.8 \times 10^4 {\rm m/s}$.}
\end{figure}

\begin{figure}
\vspace*{3.5cm}
\centerline{\epsfxsize=2in
     \epsffile{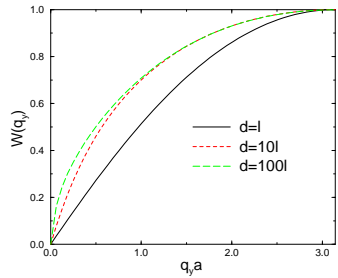}}
%\special{psfile=wghtN2.eps}
\caption[]{\label{wghtN2}
Weight function {\it vs.} $q_{y} a$ for $d=\ell$, $d = 10 \ell$ and $d = 100 \ell$.
These calculations are for $N=2$ and CDW period $a = 5.8 \ell$. 
}
\end{figure}

We now examine the large $d$, small $q$ behavior of the numerator and denominator of 
Eq.(~\ref{wght}).  For the numerator 
\begin{equation}
D_{++}(q)+{\rm Re}D_{+-}(q) = \frac{1}{4 \pi^2 \ell^2} U(q) (1 - cos(qa/2)) 
\sim \frac{e^2}{2 \pi \ell^2}  q^2 d a^{2}/2,   
\label{longrangenum}
\end{equation}
and for the denominator
\begin{equation}
D_{++}(q)^2 - |D_{+-}(q)|^2 \propto U(q) \propto   [d] \; [q^2 a^2].
\label{longrangedenom}
\end{equation}
The first factor is square brackets in Eq.(~\ref{longrangedenom}) comes from 
$D_{++}(q)+|D_{+-}(q)|$ for which the $j=0$ term dominates.  The second factor 
is proportional to $D_{++}(q)-|D_{+-}(q)|$ for which only odd $j$ terms survive,
implying no dependence on $d$ and analytic dependence on $q$.  These formulas 
apply for $q d < 1$; for $d \to \infty$, the small $q$ behavior is obtained by 
replacing $U(q=0) = 4 \pi e^2 d$ by $U(q \to 0) = 2 \pi e^2/q$, {\it i.e.}, 
by replacing $d$ by $1/2q$.  We plot the square of the denominator, 
proportional to the square of the collective excitation velocity for 
several values of the screening length $d$ in Fig.~[\ref{v2}].  The velocity
increases as $d$ increases as expected.  For $d = 100 \ell$,
the quadratic small $q$ behavior predicted in Eq.(~\ref{longrangedenom}), 
applies only for $q a  \lesssim 0.02 $, and is not apparent in the plot.
Instead we see the long range interaction behavior, in which the velocity
is proportional to $q^{1/2}$.
As is apparent from Eqs.(~\ref{noexchange}), screening is irrelevant except
very close to $q =0$ when $d$ is large.  Fig.~[\ref{v2}] shows that once 
$d \gtrsim a$, the large $d$ no-screening limit is approached.
In Fig.~[\ref{wghtN2}] we show the weight functions for $d=\ell$, 
$d=10 \ell$ and $d =100 \ell$.  At each value of $d$, the denominator 
of the weight function at small $q$ is proportional to $d^{1/2} |q|$ and 
the numerator proportional to $ d q^2$.  The weight function is therefore
proportional to $|q|$ with a coefficient which varies as $d^{1/2}$.  
A larger value of $d$ (less screening), leads to a weight function which 
increases more rapidly with $q$ and a scaling dimension for the scattering
vertex which is closer to one.  In the limit of unscreened interactions,
which applies down to small $q$ for $d = 100 \ell$, $W(q) \propto |q|^{1/2}$

\begin{figure}
\vspace*{3.5cm}
\centerline{\epsfxsize=2in
     \epsffile{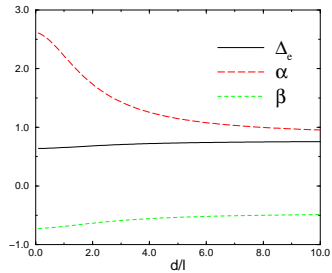}}
%\special{psfile=nltransN2.eps}
\caption[]{\label{nltrans}
Scaling dimension $\Delta_{e}$ and non-linear transport exponents 
$\alpha$ and $\beta$ for $d=\ell$, $d = 10 \ell$ and $d = 100 \ell$.  These 
calculations are for $N=2$ and CDW period $a = 5.8 \ell$.}
\end{figure}

In Fig.~[\ref{nltrans}] we show the dependence of the 
scaling dimension $\Delta_{e}$ and the non-linear transport exponents 
$\alpha$ and $\beta$ on the distance $d$ to the screening plan.
For $d \to 0$ the numerical result is very close to that from the analytic 
short-range interaction model described above which leads to 
$\Delta_{e} = 2/\pi$.  For $d \to 0$, the scaling dimension approaches 
$\Delta_{e} = 0.772$, a value we have been able to obtain only numerically.
These relatively modest changes in the scaling dimension translate into relatively
large changes in the transport exponents, particularly in $\alpha$ which 
characterizes the hard-axis non-linearity.  We predict that these non-linearities
will be much stronger if a screening placed in close proximity to the electron
layer.

\section{Summary}

Recent experiments \cite{lilly,du,shayegan} have established a consistent set of
transport properties for high-mobility two-dimensional electron systems with
high orbital index ($ N \ge 2 $) partially filled Landau levels.  These
properties differ qualitatively from those which occur 
in the low orbital index ($ N \le 1 $) fractional quantum Hall effect 
regime. At large $N$, the dissipative resistivities are large, strongly
anisotropic, and non-linear for $0.4 \lesssim \nu -
\lbrack \nu \rbrack \lesssim 0.6$ within
each Landau level. This anisotropic transport regime is bracketed by regions of
reentrant integer quantum Hall plateaus.  We have recently \cite{usprb} developed 
a theory which is able to account for most features of these experiments. 
An important prediction of the theory is that the easy and hard direction
resistivities should have non-integral power-law temperature dependences.
In this article we have briefly summarized the theory and elaborated on its
predictions for the dependence of these exponents on the distance $d$ between
the two-dimensional electron layer and a remote screening plane, predictions
which are summarized in Fig.~[\ref{nltrans}].
We find that $\Delta_{e}$ approaches two different values, both smaller than one,
for $d \to 0$ and $d \to \infty$, and interpolates smoothly between these limits at 
finite values of $d$.  Verification of our prediction that 
non-linear transport can be enhanced by introducing a screening plane and 
reducing $d$, would help substantiate our theory of quantum Hall smectics.

\section{Acknowledgements}

We would like to acknowledge insightful conversations with Herb Fertig, Michel
Fogler, Tomas Jungwirth, Jim Eisenstein, Eduardo Fradkin, Steve Girvin, and 
Philip Phillips.  AHM acknowledges the hospitality of the ITP at UC Santa Barbara,
where this work was initiated.  M.P.A.F. is grateful to the
NSF for generous support under grants DMR-97-04005, DMR95-28578 and PHY94-07194.
A.H.M. is grateful for support under grant DMR-97-14055.

\end{document}